\documentclass[12pt,preprint]{aastex}
\usepackage{amssymb}
\usepackage{amsmath}
\usepackage{graphicx}

\input tcilatex

\begin{document}

\title{Charged black holes: Wave equations for gravitational and
electromagnetic perturbations}
\author{Zolt{\'{a}}n Perj{\'{e}}s and M\'{a}ty\'{a}s Vas\'{u}th \\
KFKI Research Institute for Particle and Nuclear Physics,\\
H--1525, Budapest 114, P.O.B.\ 49, Hungary}

\begin{abstract}
A pair of wave equations for the electromagnetic and gravitational
perturbations of the charged Kerr black hole are derived. The perturbed
Einstein-Maxwell equations in a new gauge are employed in the derivation.
The wave equations refer to the perturbed Maxwell spinor $\Phi_{0}$ and to
the shear $\sigma$ of a principal null direction of the Weyl curvature. The
whole construction rests on the tripod of three distinct derivatives of the
first curvature $\kappa$ of a principal null direction.
\end{abstract}

\maketitle



\section{ Introduction}

One of the most fundamental issues of theoretical astrophysics concerns the
relativistic sources of radiation involving black holes. The basic
properties of black holes (their geometry, unicity and thermodynamical
properties) have been discovered in the middle of the twentieth century.
Early investigations by Petterson \cite{Petterson}, Chitre and Vishveshwara 
\cite{CV} indicate that the minimum energy configuration of a black hole in
an external electromagnetic field is attained when the hole has an electric
charge.

The recently discovered high-luminosity sources of X-ray radiation %
\citep{Bagaroff, Vazquez} and the need for templates of gravitational
radiation wave forms have amplified interest \citep{Hughes} in models of
black-hole perturbations \citep{Wagoner}. It has been suggested %
\citep{Ruffini,Punsly} that the observed features of gamma-ray bursters
(GRB) can be modelled by the presence of an electrically charged black hole.
There is now an abundant literature of this subject. Preparata and his
collaborators \cite{Preparata} present a model of GRB 971214.

The relativistic description of radiative processes follows two distinct
threads: the post-Newtonian expansion and black-hole perturbation theory.
Despite the recent advances in the post-Newtonian treatment of gravitational
radiation processes, it remains a laborious apparatus to work with. In
situations involving a black hole, it would seem unreasonable to perturb the
flat Minkowski background when the black hole metric is known to full
precision. The first investigation of black hole perturbations has been
undertaken by \cite{Chrzanowski}, \cite{Detweiler} and \cite{Chandrasekhar}.
In the focus of their descriptions is the decoupled equation %
\citep{Teukolsky} for the perturbed curvature quantity $\Psi_{0}$. The
ordinary differential equations resulting from the separation of this
equation are of second order, with a rugged singularity structure.

Among these pioneering studies, Chandrasekhar's own, reviewed in Sec. 9 of
his monograph \citep{Chandrasekhar}, stands out as most detailed. In a 
\textit{tour de force} of several hundred computation pages, Chandrasekhar
manages to solve his seventy-six perturbation equations. He adds, however,
the conjecture that perhaps at a later time the complexity of the problem
will be unravelled by deeper insight.

Chandrasekhar uses a \cite{NP} (NP) approach both to the unperturbed Kerr
metric and to the perturbed space-time. The null tetrad vectors of the
latter are linear combinations of those for the black-hole background. The
combination coefficients plus the Weyl tensor perturbations are fifty real
unknown functions.

The study of electromagnetic black hole perturbations has begun with the
work of Zerilli\cite{Zerilli} who considered the nonrotating system. Some
years later, the problem of charged Kerr black hole perturbations has been
taken up by Fackerell \cite{Fackerell} and Crossman \cite{Crossman}.

The purpose of the present work is precisely to help unravel the mystery of
black-hole perturbations.

In treatments of black-hole perturbations, it is a frequent practice to seek
a suitable gauge fixing. The commonly applied gauge uses the \cite%
{Kinnersley} tetrad \citep{Chitre}. Another possible choice, promoted by 
\cite{Chrzanowski2}, is the incoming (or outgoing) radiation gauge for the
normal modes. This choice does not uniquely fix the coordinate gauge,
however. In fact, the present work has been launched with the intent to take
a second look at the remaining gauge freedom. We are lead to use, however a
tetrad gauge related to the timelike Killing vector $K.$

In \cite{Perjes}, a simple pilot computation has revealed for stationary
perturbations of the charged Kerr black hole that it is possible to derive a
pair of wave equations for the electromagnetic and for the gravitational
fields in a gauge related to the timelike Killing vector $K=\partial
/\partial t$. We now show that this result can be extended to arbitrary
perturbations when using the \cite{Newman} tetrad. To our delight, the two
wave equations governing the Maxwell field function $\Phi _{0}$ and the
shear $\sigma $ of a principal null direction of the Weyl tensor survive the
generalization to arbitrary perturbations! A brief account of this latter
result has been published in \cite{Perjes2}. The present paper discusses the
full details.

In Section 2, we collect the relations describing the Kerr-Newman solution
of the coupled Einstein-Maxwell equations. In section 3, we consider the
problem of choosing the tetrad for the perturbed space-time, and derive a
triple of expressions for three derivatives of the first curvature $\kappa$
of a principal null direction. The whole description of black-hole
perturbations rests on this tripod. In Section 4, the integrability
conditions of these derivatives will provide us with the key wave equations.
The boundary conditions on the horizon and at null infinity are studied in
Section 5.

\section{The unperturbed variables}

The current literature of black hole physics predominantly utilizes the
Boyer-Lindquist coordinates $\left( \hat{t},r,\vartheta ,\hat{\varphi}%
\right) $ which are related to the Kerr-Newman coordinates $\left(
t,r,\vartheta ,\varphi \right) $ by the transformation 
\begin{eqnarray}
d\hat{t} &=&-dt+\tfrac{r^2+a^2}{\mathbf{\Delta }}dr  \label{dtdphi} \\
d\hat{\varphi} &=&-d\varphi +\tfrac a{\mathbf{\Delta }}dr  \notag
\end{eqnarray}
where 
\begin{equation}
\mathbf{\Delta }=r^2-2\mathrm{m}r+a^2+\mathrm{e}^2
\end{equation}
is the horizon function. The effect of the transformation is a simultaneous
reflection of the time and lattitude directions, and a shift of the origin
of the $t$ and $\varphi $ coordinates along the open $t$ trajectories and
circular $\varphi $ trajectories, respectively, by an amount dependent on
the radial coordinate $r.$ Although the Boyer-Lindquist coordinates have the
salient feature that the two double principal null directions of the
curvature are represented symmetrically\cite{MWT}, they cover only the part
of the original $\left( t,r,\vartheta ,\varphi \right) $ coordinate domain
extending between null infinity and the horizon. (However, this symmetry is
restored in the original coordinates by using a twin transformation to Eqs. (%
\ref{dtdphi}) where the directions on the trajectories are not reversed).

In the original coordinates the charged Kerr metric is \citep{Newman} 
\begin{align}
ds^{2}& =\left( 1-\frac{2\mathrm{m}r-\mathrm{e}^{2}}{\zeta \overline{\zeta }}%
\right) \left( dt-a\sin ^{2}\vartheta d\varphi \right) ^{2}  \label{ds2} \\
& +2\left( dt-a\sin ^{2}\vartheta d\varphi \right) \left( dr+a\sin
^{2}\vartheta d\varphi \right) -\zeta \overline{\zeta }\left( d\vartheta
^{2}+\sin ^{2}\vartheta d\varphi ^{2}\right)  \notag
\end{align}%
with $\mathrm{m}$ the mass, $\mathrm{e}$ the electric charge, $\mathrm{m}a$
the angular momentum and 
\begin{equation*}
\zeta =r-ia\cos \vartheta .
\end{equation*}%
An overbar denotes complex conjugation. The corresponding null tetrad is %
\citep{Newman} 
\begin{align}
D& \equiv \ell ^{a}\frac{\partial }{\partial x^{a}}=\frac{\partial }{%
\partial r}  \notag \\
\Delta & \equiv n^{a}\frac{\partial }{\partial x^{a}}=\frac{1}{2}\left[ 
\frac{\left( \zeta +\overline{\zeta }\right) \mathrm{m}-\mathrm{e}^{2}}{%
\zeta \overline{\zeta }}-1\right] \frac{\partial }{\partial r}+\frac{%
\partial }{\partial t}  \notag \\
\delta & \equiv m^{a}\frac{\partial }{\partial x^{a}}=\frac{1}{2^{1/2}%
\overline{\zeta }}\left[ \frac{\partial }{\partial \vartheta }+\frac{i}{\sin
\vartheta }\frac{\partial }{\partial \varphi }-ia\sin \vartheta \left( \frac{%
\partial }{\partial r}-\frac{\partial }{\partial t}\right) \right] \\
\overline{\delta }& \equiv \overline{m}^{a}\frac{\partial }{\partial x^{a}}.
\notag
\end{align}
The real null vector $\ell $ of this tetrad points in one of the double
principal directions of the space-time, and the vector $n$ lies in the
intersection of the two-plane of the vector $\ell $ and the vector $K$ with
the null cone. (Note that the literature of black hole perturbations is
biased towards the use of the Kinnersley tetrad \cite{MWT} as a companion of
the Boyer-Lindquist coordinates, which we do not use in this paper). In the
NP notation, the following spin coefficients have nonzero values 
\begin{align}
\rho & =-\frac{1}{\zeta }\qquad \qquad \qquad \quad \qquad \gamma =\frac{%
\mathrm{m}\overline{\zeta }-\mathrm{e}^{2}}{2\zeta ^{2}\overline{\zeta }} 
\notag \\
\alpha & =-\frac{1}{2^{3/2}\zeta }\frac{\cos \vartheta }{\sin \vartheta }%
\qquad \quad \qquad \beta =\frac{1}{2^{3/2}\overline{\zeta }}\frac{\cos
\vartheta }{\sin \vartheta } \\
\qquad \mu & =\frac{\mathrm{m}(\zeta +\overline{\zeta })-\mathrm{e}^{2}}{%
2\zeta ^{2}\overline{\zeta }}-\frac{1}{2\overline{\zeta }}\qquad \nu =ia\sin
\vartheta \frac{\mathrm{m}\overline{\zeta }-\mathrm{e}^{2}}{2^{1/2}\zeta ^{3}%
\overline{\zeta }}.  \notag
\end{align}%
The nonvanishing components of the Weyl spinor are 
\begin{align}
\Psi _{2}& =\frac{\mathrm{e}^{2}-\mathrm{m}\overline{\zeta }}{\zeta ^{3}%
\overline{\zeta }}  \notag \\
\Psi _{3}& =-3ia\sin \vartheta \frac{\mathrm{m}\overline{\zeta }-\mathrm{e}%
^{2}}{2^{1/2}\zeta ^{4}\overline{\zeta }}  \label{Psikerr} \\
\Psi _{4}& =3a^{2}\sin ^{2}\vartheta \frac{\mathrm{m}\overline{\zeta }-%
\mathrm{e}^{2}}{\zeta ^{5}\overline{\zeta }}  \notag
\end{align}%
and the nonzero Maxwell spinor components have the form 
\begin{align}
\Phi _{1}& \equiv \frac{1}{2}F_{ab}\left( \ell ^{a}n^{b}+\overline{m}%
^{a}m^{b}\right) =\frac{\mathrm{e}}{2^{1/2}\zeta ^{2}}  \notag \\
\Phi _{2}& \equiv F_{ab}\overline{m}^{a}n^{b}=\frac{i\mathrm{e}a\sin
\vartheta }{\zeta ^{3}}.
\end{align}

In the following sections, we develop a perturbative approach to the charged
Kerr black hole. We shall focus on the small quantities in the perturbed
space-time. It is, therefore, important to realize that the NP quantities $%
\sigma,\kappa,\epsilon,\lambda,\pi,\tau,\bar{\alpha}+\beta,\Psi_{0},\Psi_{1}$
and $\Phi_{0}$ all vanish in the present gauge.

\section{Choice of gauge}

In this section, we introduce a new gauge for the coupled electromagnetic
and gravitational perturbations of the charged Kerr black hole in which the
wave equations decouple, as found in the presence of a Killing symmetry %
\citep{Perjes}.

The vector $\ell $ is chosen along a principal null direction of the
perturbed space-time: 
\begin{equation}
\Psi _{0}=0.  \label{pnd}
\end{equation}%
Unlike in the case of the principal null spinor of the exact Weyl tensor,
this condition does not determine uniquely the spinor $o^{A}$. Infinitesimal
spinor rotations of the form 
\begin{equation}
o^{A}\rightarrow o^{A}+b\iota ^{A},\qquad \iota ^{A}\rightarrow \iota ^{A},
\label{dyadtr}
\end{equation}%
where $b$ is an arbitrary first-order multiplying function, are allowed. The
tetrad and spin coefficient transformations induced by these rotations are
given in Eqs. (II) and (343) of Ref. \cite{Chandrasekhar}, respectively.

The gauge condition (\ref{pnd}) is preserved by the dyad transformation (\ref%
{dyadtr}). In this gauge, the electromagnetic Stewart-Walker identity %
\citep{Stewart} 
\begin{align}
& [(D\!-\!\epsilon \!+\!\bar{\epsilon}\!-\!2\rho \!-\!\bar{\rho})(\Delta
\!+\!\mu \!-\!2\gamma )\!-\!(\delta \!-\!\beta \!-\!\bar{\alpha}\!-\!2\tau
\!+\!\bar{\pi})(\bar{\delta}\!+\!\pi \!-\!2\alpha )\!+\!\sigma \lambda
\!-\!\kappa \nu ]\Phi _{0}=  \notag \\
\!\!\!\!\!\!& -2[\kappa \left( \Delta -\!3\gamma \!-\!\bar{\gamma}\!+\!\bar{%
\mu}\right) \!-\!\sigma \left( \bar{\delta}\!-\!3\alpha \!+\bar{\beta}\!-%
\bar{\tau}\!\right) +\!\Delta \kappa \!-\!\bar{\delta}\sigma \!+\!2\Psi
_{1}]\Phi _{1}\!+\!\Psi _{0}\Phi _{2}.  \label{EMSW}
\end{align}%
takes a simpler form since the last term on the right vanishes. The function 
$b$ is then utilised, following Crossman's idea \citep{Crossman}, to make
the whole right-hand side vanish. The latter condition is employed to
express the derivative 
\begin{equation}
\Delta \kappa =\left( \bar{\delta}-3\alpha +\bar{\beta}-\bar{\tau}+\frac{%
\bar{\delta}\Phi _{1}}{\Phi _{1}}\right) \sigma +\left( 3\gamma +\bar{\gamma}%
-\bar{\mu}-\frac{\Delta \Phi _{1}}{\Phi _{1}}\right) \kappa -2\Psi _{1}.
\label{Delka}
\end{equation}

In what follows, we use a linear approximation, that is to say, we neglect
products and higher powers of quantities which vanish for the Kerr-Newman
space-time. The operators in the parentheses in Eq. (\ref{Delka}) act on the
first-order functions $\sigma$ and $\kappa$, hence we keep their form for
the charged Kerr metric. For example, $\Phi_{1}=\mathrm{e}/\left(
2^{1/2}\zeta^{2}\right) ,$ where the numerator (the electric charge $\mathrm{%
e}$) exactly cancels in the bracketed terms containing the field $\Phi_{1}.$
Thus we may keep our gauge condition (\ref{Delka}) in the limiting case of
an uncharged black hole, \textit{i.e.}, when $\mathrm{e}=0,$ even though
nothing remains of the original motivation for adopting it. In
Chandrasekhar's terms \citep{Chandrasekhar}, the gauge (\ref{Delka}) becomes
a phantom gauge.

Throughout, we express the derivatives $\delta\rho,D\rho,\Delta\rho ,D\bar{%
\alpha},D\tau,D\mu$ and $D\gamma$ from the NP equations (4.2k), (4.2a),
(4.2q), (4.2d), (4.2c), (4.2h) and (4.2f), respectively; $D\Phi_{2}$ and $%
\delta\Phi_{2}$ from Maxwell's equations (A1) and $D\Psi_{2}$ and $\delta
\Psi_{2}$ from the Bianchi identities (A3) \citep{NP}.

A relation for $D\kappa$ can be obtained from the NP commutator \citep{NP} 
\begin{equation}
\delta D-D\delta=\left( \bar{\alpha}+\beta-\bar{\pi}\right) D+\kappa
\Delta-\sigma\bar{\delta}-\left( \bar{\rho}+\epsilon-\bar{\epsilon}\right)
\delta  \label{comm2}
\end{equation}
acting on $\Psi_{1}$, after expressing both $D\Psi_{1}$ and $\delta\Psi_{1}$
from the Bianchi identities. This has the form 
\begin{align}
D\kappa & =2\frac{\bar{\Phi}_{1}}{\bar{\Phi}_{2}}D\sigma-\kappa\rho-\frac {1%
}{2\Phi_{1}}\ \left( D-4\rho\right) D\Phi_{0}  \label{Dka} \\
& +\frac{\bar{\Phi}_{1}}{2\bar{\Phi}_{2}\Phi_{1}}\left[ \delta D+D\delta+(%
\bar{\rho}-4\rho)\delta+4\bar{\alpha}(D+\bar{\rho}-2\rho)\right] \Phi_{0}. 
\notag
\end{align}
A third expression for a derivative of the spin coefficient $\kappa$ is
available from the NP equation (4.2b): 
\begin{equation}
\delta\kappa=D\sigma-\left( \rho+\bar{\rho}\right) \sigma+\left( \bar{\alpha}%
+3\beta\right) \kappa.  \label{deka}
\end{equation}

The integrability conditions of the derivatives $\Delta\kappa$, $D\kappa$
and $\delta\kappa$ will be explored in the next section.

\section{\protect\bigskip Master equations\label{bhpert4r.red}}

In this section, we continue to use the gauge (\ref{Delka}) in which Eq. (%
\ref{EMSW}) takes the form 
\begin{align}
& [\delta \bar{\delta}-D\Delta +(2\gamma -\mu )D-2\alpha \delta +(\bar{\rho}%
+2\rho )\Delta  \label{Phieq} \\
& +2\mu \rho -2\gamma \left( 2\rho +\bar{\rho}\right) -2\delta \alpha +\Psi
_2+2\bar{\Phi}_1\Phi _1]\Phi _0=0.  \notag
\end{align}
Inserting the unperturbed quantities in Eq. (\ref{Phieq}), we get the first
wave equation: 
\begin{equation}
\square _1\Phi _{\mathbf{0}}=0,  \label{ip1}
\end{equation}
where we introduce the operator 
\begin{align}
\square _s& =\mathbf{\Delta }^{-s}\frac \partial {\partial r}\mathbf{\Delta }%
^{s+1}\frac \partial {\partial r}+\frac 1{\sin \vartheta }\frac \partial
{\partial \vartheta }\sin \vartheta \frac \partial {\partial \vartheta
}+s\left( 1-s\frac{\cos ^2\vartheta }{\sin ^2\vartheta }\right)  \notag \\
& \mathcal{+}\left[ 2a\left( \frac \partial {\partial t}-\frac \partial
{\partial r}\right) +\frac 1{\sin ^2\vartheta }\left( \frac \partial
{\partial \varphi }+2is\cos \vartheta \right) \right] \frac \partial
{\partial \varphi } \\
& +a^2\sin ^2\vartheta \frac{\partial ^2}{\partial t^2}-2\left[
(r^2+a^2)\frac \partial {\partial r}+(s+2)r+ia\cos \vartheta \right] \frac
\partial {\partial t}.  \notag
\end{align}
The spectral condition $\square _s\Phi _{\QTR{textbf}{0}}=\chi \Phi _{%
\QTR{textbf}{0}}$ is separable in the coordinates $r$ and $\vartheta $ on
the basis of mode functions of the form $f\left( r,\vartheta \right)
e^{i(m\varphi -\omega t)}$. The homogeneous wave equation (\ref{ip1}) can be
brought to the form 
\begin{equation}
\left[ \left( \nabla ^a+s\Gamma ^a\right) \left( \nabla _a+s\Gamma _a\right)
-4s^2\Psi _2\right] \Phi _0=0
\end{equation}
with $\Psi _2$ is as given in Eq.(\ref{Psikerr}), $s=1$ and 
\begin{equation}
\Gamma ^r=\tfrac 2\zeta ,\qquad \Gamma ^\vartheta =0,\qquad \Gamma ^\varphi
=i\cos \vartheta +a\sin ^2\vartheta \tfrac{\zeta \overline{\zeta }+\mathrm{m}%
(2\overline{\zeta }+\zeta )-2\mathrm{e}^2}{\zeta ^2\overline{\zeta }},\qquad
\Gamma ^t=\tfrac{\zeta \overline{\zeta }-\mathrm{m}(2\overline{\zeta }+\zeta
)+2\mathrm{e}^2}{\zeta ^2\overline{\zeta }}.
\end{equation}
Equations in this class have been studied by Bini, Cherubini, Jantzen and
Ruffini\cite{Bini}.

The second wave equation arises as follows. Given the three different
derivatives (\ref{Delka}), (\ref{Dka}) and (\ref{deka}) of $\kappa,$ we can
derive three equations from the commutation relations. Only one of these
equations is a second order differential equation decoupled from the other
variables. It is obtained by applying the NP commutator 
\begin{equation}
\delta\Delta-\Delta\delta=-\overline{\nu}D+\left( \tau-\bar{\alpha}%
-\beta\right) \Delta-\overline{\lambda}\bar{\delta}-\left( \mu -\gamma+%
\overline{\gamma}\right) \delta  \label{comm3}
\end{equation}
to $\kappa,$ and substituting the second derivatives $\delta\Delta\Phi_{1}$
and $\delta\bar{\delta}\Phi_{1}$ from the NP commutators (\ref{comm3}) and 
\begin{equation}
\bar{\delta}\delta-\delta\bar{\delta}=\left( \overline{\mu}-\mu\right)
D+\left( \overline{\rho}-\rho\right) \Delta-\left( \bar{\alpha}-\beta\right) 
\bar{\delta}-\left( \overline{\beta}-\alpha\right) \delta  \label{comm4}
\end{equation}
for $\Phi_{1}$. The resulting equation has the form 
\begin{equation}
\mathcal{D}_{1}\sigma+\mathcal{D}_{2}\Phi_{0}+\mathcal{A}\kappa=0,
\label{T2abs}
\end{equation}
where

\begin{align}
\mathcal{D}_{1} & =2\{\bar{\Phi}_{2}\Phi_{1}[\delta\bar{\delta}-\Delta
D-4\alpha\delta+2\bar{\alpha}\bar{\delta}+(\bar{\rho}+\rho)\Delta  \notag \\
& +\bar{\gamma}\rho+4\bar{\Phi}_{1}\Phi_{1}-8\alpha\bar{\alpha}+\Delta \bar{%
\rho}-4\delta\alpha+(\bar{\rho}-2\rho)\bar{\mu}  \notag \\
& +3\left( \mu-\gamma\right) \rho-(4\gamma-\mu)\bar{\rho}+5\Psi _{2}]  \notag
\\
& -\Phi_{1}\left[ \bar{\mu}\bar{\Phi}_{2}-2\bar{\nu}\bar{\Phi}%
_{1}-(4\gamma-\mu)\bar{\Phi}_{2}\right] D  \notag \\
& +\bar{\Phi}_{2}\left[ \bar{\delta}\Phi_{1}(\delta+4\bar{\alpha}%
)-\Delta\Phi_{1}\left( D-2\rho\right) \right] \}  \notag \\
\mathcal{D}_{2} & =\bar{\nu}\left[ \bar{\Phi}_{1}(D\delta+\delta D)-\bar{\Phi%
}_{2}DD\right] +\left[ (\bar{\rho}-4\rho)\bar{\nu}+4\bar{\Phi }_{2}\Phi_{1}%
\right] \bar{\Phi}_{1}\delta  \notag \\
& +4\left[ (\bar{\nu}\rho-\bar{\Phi}_{2}\Phi_{1})\bar{\Phi}_{2}+\bar{\alpha }%
\bar{\nu}\bar{\Phi}_{1}\right] D+4\left[ (\bar{\rho}-2\rho)\bar{\nu}+2\bar{%
\Phi}_{2}\Phi_{1}\right] \bar{\alpha}\bar{\Phi}_{1}  \notag \\
\mathcal{A} & =2\bar{\Phi}_{2}\Phi_{1}\left[ \delta\left( 3\gamma +\bar{%
\gamma}-\bar{\mu}\right) +2(\Delta+\mu+\bar{\gamma}-\gamma)\bar{\alpha }+%
\bar{\nu}\rho-4\bar{\Phi}_{2}\Phi_{1}\right] .
\end{align}
The operators $\mathcal{D}_{1},$ $\mathcal{D}_{2}$ and $\mathcal{A}$ can be
taken to have their form in the Kerr-Newman space-time since they act on
first-order functions. The term $\mathcal{A}\kappa$ in Eq. (\ref{T2abs})
vanishes.

It proves advantageous to introduce in (\ref{T2abs}) the `\textit{news
potential'} $\psi $ by 
\begin{equation}
\psi =\frac{\sigma }{\zeta ^{2}}.  \label{psidef}
\end{equation}%
We then have the wave equation of the form 
\begin{equation}
\square _{2}\psi =\frac{1}{\overline{\zeta }^{2}\mathrm{e}}J\Phi _{0},
\label{psieq}
\end{equation}%
where 
\begin{align}
J& =\left( \mathrm{e}^{2}-\mathrm{m}\zeta \right) \left( \frac{\partial }{%
\partial \vartheta }+ia\sin \vartheta \frac{\partial }{\partial t}+\frac{i}{%
\sin \vartheta }\frac{\partial }{\partial \varphi }-\frac{\cos \vartheta }{%
\sin \vartheta }\right) \frac{\partial }{\partial r}  \notag \\
& -\frac{1}{\overline{\zeta }}\left[ \mathrm{e}^{2}+\mathrm{m}\left( 2%
\overline{\zeta }-\zeta \right) \right] \left( \frac{\partial }{\partial
\vartheta }+ia\sin \vartheta \frac{\partial }{\partial t}+\frac{i}{\sin
\vartheta }\frac{\partial }{\partial \varphi }-\frac{\cos \vartheta }{\sin
\vartheta }\right)  \\
& +\frac{1}{\overline{\zeta }}\left[ \mathrm{e}^{2}-\mathrm{m}\left( 2%
\overline{\zeta }+\zeta \right) \right] ia\sin \vartheta \frac{\partial }{%
\partial r}.  \notag
\end{align}%
Both equations (\ref{Phieq}) and (\ref{psieq}) pick up an additional source
term in the presence of other form of matter, and these terms contain the
corresponding stress-energy tensor $T_{ab}.$ Following \cite{Teukolsky}, we
expand the homogeneous solutions in quasi-normal modes with energy $\omega $
and helicity $m$ : 
\begin{equation}
\Phi _{0}=\int d\omega \sum_{l,m}R\left( r\right) S_{l}^{m}\left( \vartheta
\right) e^{i(m\varphi -\omega t)}.
\end{equation}%
Separating the kernel of the operator $\square _{s}$, the radial function $%
R\left( r\right) $ and the angular function $S_{l}^{m}\left( \vartheta
\right) $ satisfy the ordinary differential equations, respectively, 
\begin{align}
\!\!\!\!\!\!\!\!\!\left[ \mathbf{\Delta }^{-s}\frac{\partial }{\partial r}%
\mathbf{\Delta }^{s+1}\frac{\partial }{\partial r}\!+\!2i\left[
(r^{2}\!+\!a^{2})\omega \!-\!am\right] \frac{\partial }{\partial r}%
\!+\!2i\omega (s\!+\!2)r\!+\!2am\omega \!-\!\Lambda \right] R\!\!\!\!\!&
=\!\!\!\!\!0  \label{radeq} \\
\!\!\!\!\!\left[ \frac{1}{\sin \vartheta }\frac{\partial }{\partial
\vartheta }\sin \vartheta \frac{\partial }{\partial \vartheta }\mathcal{-}%
\!\left( \frac{m+s\cos \vartheta }{\sin \vartheta }\right) ^{2}\!+\!\left(
\omega a\cos \vartheta \!-\!1\right) ^{2}\!+\!s\!-\!1\!+\!\Lambda
\!-\!\omega ^{2}a^{2}\right] S\!\!\!\!\!& =\!\!\!\!\!0  \label{angeq}
\end{align}%
and $\Lambda $ is the separation constant. Solution techniques for equations
of the type (\ref{radeq}) and (\ref{angeq}) have been developed in \cite%
{Mano,Mano2}.

\section{Boundary conditions}

At the horizon, $\mathbf{\Delta }=0,$ the radial equation (\ref{radeq}) is
singular, even though the metric (\ref{ds2}) remains here regular. We shall
impose the boundary conditions at null infinity, $r\rightarrow \infty $ and
on the horizon. To this end, we perform a transformation of both the
dependent variable and the radial coordinate: 
\begin{equation}
Z=e^{iA}\mathbf{\Delta }^{s/2}(r^{2}+a^{2})^{1/2}R\ ,\qquad \frac{\partial
r^{\prime }}{\partial r}=\frac{r^{2}+a^{2}}{\mathbf{\Delta }}\mathbf{,}
\end{equation}%
where $Z$ and $r^{\prime }$ are the new radial function and coordinate,
respectively. We choose the real function $A$ as follows, 
\begin{equation}
\mathbf{\Delta }A_{,r}=(r^{2}+a^{2})\omega -am.  \label{A}
\end{equation}%
As a result, the first derivatives of $Z$ do not appear in the radial
equation 
\begin{equation}
Z_{,r^{\prime }r^{\prime }}+UZ=0
\end{equation}%
with 
\begin{align}
U& =\frac{Wr\left[ 3Wr+2\left( \mathrm{m}-r\right) \right] -V^{2}}{%
(r^{2}+a^{2})^{2}}  \notag \\
& {+}\left[ {2(s+1)ir\omega +2am\omega -\Lambda -}s-W\right] \frac{{W}}{%
r^{2}+a^{2}} \\
V& =(r-\mathrm{m})s+i\left[ {(r^{2}+a}^{2})\omega -am\right]  \\
W& =\frac{\mathbf{\Delta }}{r^{2}+a^{2}}.
\end{align}

(i) In the neighborhood of null infinity, for large values of $r,$ we expand
the equation (\ref{radeq}) in powers of $1/r:$

\begin{equation}
Z_{,r^{\prime}r^{\prime}}+\left( \omega^{2}+\frac{2i\omega}{r}\right)
Z\approx0.  \label{Zas}
\end{equation}
The asymptotic solutions of (\ref{Zas}) and (\ref{A}) are $Z\sim r^{\pm
1}e^{\mp i\omega r^{\prime}}$ and the asymptotic form of $A$ is $A\sim\omega
r$. Hence the asymptotic form of the radial function is $R\sim
r^{-s}e^{-2i\omega r^{\prime}}$ and $r^{-s-2}$.

(ii) At the event horizon, $r=r_{+}$ or $r^{\prime}\rightarrow-\infty$, the
radial equation becomes 
\begin{equation}
Z_{,r^{\prime}r^{\prime}}-\frac{V^{2}}{(a^{2}+r_{+}^{2})^{2}}Z\approx0.
\end{equation}

The boundary conditions on the event horizon will be selected by taking into
account the absence or presence of incoming or outgoing radiation.

The boundary condition for the angular equation (\ref{angeq}), imposed at $%
\vartheta=0$ and at $\vartheta=\pi$ is that $S$ must be regular on the
boundary. As before \citep{Teukolsky}, this leads to the Sturm-Liouville
eigenvalue problem and yields a complete set of complex solutions.


\section{Ramifications}

The Maxwell equations for electromagnetic perturbations on a fixed
space-time are homogeneous. The circularly polarized normal modes have the
structure $\phi_{A}\left( r,\vartheta\right) e^{i\left( m\varphi-\omega
t\right) }$. The interaction with the gravitational field, via the
stress-energy tensor $\phi_{A}\overline{\phi}_{B},$ introduces a mixing
among the electromagnetic modes, even in the linearized theory. Thus, the
full electrovacuum perturbation equations cease to be homogeneous in the
modes. In quantum field theory, perturbative methods are available for such
interacting systems. Strangely, no comparable treatment for the
corresponding classical system is known. This regrettable backlog in the
classical theory, however, will help illuminate the origin of the
long-standing difficulties in finding a separable equation for the
perturbations.

In this paper, the following picture emerges for the classical electrovacuum
perturbations. There exists a \emph{subset} of the field equations,
consisting, in part, of gravitational equations and of some of the Maxwell
equations not containing any mode mixing. (This is because the complex
conjugate electromagnetic stresses are absent from these equations). From
these relations alone, it is possible to obtain a \emph{pair} of wave
equations for the quantities $\Phi_{0}$ and $\sigma.$ Thus a normal mode
expansion for this doublet of fields \QTR{textit}{is} available.

Given the perturbed quantities $\Phi _{0}$ and $\sigma $, Eq. (\ref{Dka}) is
an inhomogeneous first-order ordinary differential equation for the function 
$\kappa $. Integration of Eq. (\ref{Dka}) yields the $r$ dependence of the
spin coefficient $\kappa $. Once this is known, the $r$ dependence of
expressions like $\Delta \kappa $ is fixed. We then get \ $\Psi _{1}$ by
solving the linear algebraic Eq. (\ref{Delka}) for $\Psi _{1}$. Mode mixing
occurs only in the further gravitational and electromagnetic perturbation
components.

$\qquad$

\section{Acknowledgments}

This work has been supported by the OTKA grant T031724.

\end{document}